\documentclass[format=acmsmall, review=false]{acmart}

\usepackage{acm-ec-19}
\begin{document}
\title{Data Consortia}

\author{Eric Bax, John Donald, Melissa Gerber, Lisa Giaffo, Tanisha Sharma, Nikki Thompson, Kimberly Williams (Future Studies Group, Verizon Media)\\
(email: ebax@verizonmedia.com)}
%\authornote{}
%\affiliation{\institution{Future Studies Group, Verizon Media}}
%\email{baxhome@yahoo.com}

\renewcommand{\shortauthors}{FSGVM}

\begin{abstract}
Today, web-based companies use user data to provide and enhance services to users, both individually and collectively. Some also analyze user data for other purposes, for example to select advertisements or price offers for users. Some even use or allow the data to be used to evaluate investments in financial markets. Users' concerns about how their data is or may be used has prompted legislative action in the European Union and congressional questioning in the United States. But data can also benefit society, for example giving early warnings for disease outbreaks, allowing in-depth study of relationships between genetics and disease, and elucidating local and macroeconomic trends in a timely manner. So, instead of just a focus on privacy, in the future, users may insist that their data be used on their behalf. We explore potential frameworks for groups of consenting, informed users to pool their data for their own benefit and that of society, discussing directions, challenges, and evolution for such efforts.
\end{abstract}

\keywords{data, privacy, prediction, society, policy}

\maketitle

\section{Introduction}
Web-based organizations access, store, and analyze user data in ways that enhance their users' lives. Users see more relevant search results more quickly because their past searches are used to determine which results are most likely of interest to them and, collectively, to offer a selection of query completions, saving the need to type long, exact queries. Email users benefit from having their emails stored, organized, and indexed for quick search. They also benefit from email providers analyzing patterns across emails to determine which emails are spam or contain links to malware. Users of online media services benefit from collaborative filtering over user viewing and listening data to provide better recommendations about what to experience next. In each case, users allow organizations to access data about the users in order to provide the users with better service.

Web-based organizations also analyze user data to monetize their services. Many web-based services are offered to users without direct cost. In exchange, users experience advertisements, which are selected based on user data. Done well, targeted advertising means that users experience relevant messages about products, brands, and issues that interest them, advertisers connect with interested users, and the quality of advertising reflects positively on the media itself. Done poorly, it can leave users with the impression that their attention has been abused and that their data has been used against them, enabling uninteresting, obnoxious, or offensive messages to chase users around the web. The same can occur with paid services as well, offending users with efforts to elicit future payments, to elicit feedback on their services or providers, or to recruit users as an informal salesforce targeting their friends.

Concerns about these and other issues around the use of data have prompted legislative action in the European Union, in the form of the General Data Protection Regulation (GDPR) \cite{gdpr16}, and legislative interest in the United States, most notably in the form of congressional hearings.  Tim Cook, CEO of Apple, has spoken in favor of regulations that allow users to have more knowledge of and control over how their data is used \cite{cook18}. Users have begun to ask questions like "What is being done with my data without my knowledge?" and "Is my data being used against me?" Users are beginning to understand that their data has value. In the future, in the same way that people expect a return on money, we predict that users will expect a return on the aggregation and analysis of their data.

\section{Data Consortia}
Our goal is to explore the potential for groups of people to pool and invest their data for individual or societal returns. We use the term \textit{data consortium} to mean an organization with members and an administration team, in which

\begin{itemize}
\item Data consortium members grant data access to a consortium's administration team. Each member selects which data sources it allows the consortium to access, and which filters to place on data collected by the consortium. For example, a user may select to share their email inbox and their browsing behavior with the consortium, but not records of which shows they watch on their television. And a user may agree to grant access to their email inbox, but only for receipts, not for personal emails, and only for the amounts of receipts, items purchased, and company sending the receipt. 

\item The consortium's administration team establishes access to the data it is permitted to collect. This step may include receiving external access to the data as if a user or making an agreement with a data collecting entity, such as an email provider or web browser supplier, to receive the data.

\item The consortium's administration team uses computer systems to aggregate the user data and analyze it. The analysis may include trends, analysis by locale and demographics,  weighting data for different users to make the consortium data a better reflection of some other population, and identifying sets of consortium members whose data most effectively predicts different types of trends. One example of analysis is to compute aggregated trend data about consortium members' purchases from different companies, indicating which companies are increasing their revenue and which are decreasing.

\item The consortium's administration team may sell some analysis, for example to investment companies, and pay some of the proceeds to consortium members. 

\item The administration team itself may use the analysis as an input to decide on which companies to invest in, returning some profits to or giving some payment to consortium members, and/or give consortium members beneficial terms as investors, and/or restrict investment based on the analysis only to consortium members. 

\item The consoritum administration team may publish the results of some analysis to the consortium members. This may include fashion-buying trends, alerts when infectious disease levels are elevated in a locale, trends in popularity of tv shows or movies by demographics, whether people are buying more books or using libraries more, and which food delivery services receive the fewest complaints and cancellations. 

\item Results of analysis may be used to automatically adjust users' data-generating experiences. For example, if analysis shows that most consortium members who buy from a company return what they have bought, then a web browser extension for consortium members may warn members who navigate to the company's web site. 

\item Payments to consortium members may be based on the amount of data they contribute and on how useful that data is for making accurate predictions, profitable decisions, or valuable insights.
\end{itemize}

The administration team may be part of a larger organization, for example a team within an organization that already hosts member data. Alternatively, the administration team may be drawn from the members and perhaps use software and processes supplied by a third party that specializes in organizing data consortia, or the administration team may be a third-party company that serves that function for one or more data consortia.

Each data consortium may use member data for a specific purpose, with members perhaps joining multiple data consortia for different purposes, each with an administration team focused on its purpose -- a type of vertical specialization. Or there may be horizontal specialization: different organizations developing expertise in different tasks, such as data extraction, aggregation, analysis and execution for different purposes, with administration teams relying on these external organizations to accomplish some of their tasks, making it easier for a single data consortium to use member data for multiple purposes.

Organizations that currently store and manage user data may offer computation and interfaces for data consortia to more conveniently access member data, charging for the service while perhaps enhancing security and lowering bandwidth requirements by doing some of the data filtering and analysis at the site of the data before transmitting it to the data consortia. Data access and expertise gives these organizations an advantage should they choose to offer the services of a data consortium to their existing users. 

The next few sections discuss in more detail some potential use cases in a few areas: financial investment, consumer spending, and informed living. Then we conclude with a discussion of potential challenges for data consortia.

\section{Financial Investment}
Aggregated user-generated data can provide accurate and actionable financial forecasts and measurements. Examples include economic forecasts based on search data \cite{hellerstein12,damuri17,onder15} and equity investment decisions \cite{quandl_email_receipts,quandl_email_gopro,alt_data} and market insights \cite{kooti17,kooti17b} based on analyzing email receipts.

Data-based investment requires data, analysis, and capital. In the most encapsulated form, data consortium members could supply all three, forming an online version of an investment club. Alternatively, members could supply data and capital, and arrange for a financial institution to supply analysis, along with decision-making, trade execution, and accounting. A data consortium could also use outside capital, allowing external investors who could pay a fee to the consortium for use of its data. In the least encapsulated forms, a data consortium simply sells investors or investment institutions access to aggregated member data or analysis of it. 

Using data consortium member data for investment decisions raises some interesting issues and questions. One issue is inadvertent insider trading. Members' insider status would need to be recorded in order to avoid collecting data from insiders, and other members' data may also need to be checked for whether it contains insider information. This is less true just for scanning for receipts and more true for unstructured text, which may hold valuable information and may be explored in a much more automated way than receipts, making it more difficult to structurally assess what is and is not insider information. 

One question is how value and compensate for data vs. analysis and execution vs. capital. Inevitably, the answers will vary, as they do for the division of investment gains and losses today between capital and administration. For some data consortia, members may wish to be compensated based on the value of their individual data. That may seem impossible, much as micro payments once did. However, a combination of big-data analysis, and micro-economic theory can do just this. For details, refer to Appendix \ref{value_data}.

%\section{Consumer Choices}
%(***Other things -- groceries, etc. where to shop***)

%Kardashians, but with privacy
%wealthy, localities sell aggregate data -- what cars they buy and shows they watch
%people aspire to success, and when they achieve it want to know how best to enjoy it
%or even want to know how to appear successful

\section{Consumer Spending}
There are existing services designed to use individual user data for the financial advantage of users. Earny and Paribus analyze member emails for receipts, then monitor prices for items bought. If prices fall within a specified time period, then Earny and Paribus arrange partial refunds for their members. Trim analyzes emails and other personal data sources to detect subscriptions, present them to users, and cancel subscriptions on behalf of users who no longer desire them, or, in some cases, were never even aware of them. BillShark analyzes user billing data and negotiates on behalf of users to lower their bills.

Note that these services also use aggregate user data, even if indirectly. Earny, Paribus, and Trim use the fact that they act on behalf of many users to achieve economies of scale through automation, making it much easier for users to get discounts or cancel subscriptions through their services than on their own. Trim and BillShark can invest much more time and resources to learning which deals are the best available than individual consumers can, because these companies' representatives become experts through experience and because they can amortize the cost of research over their customer bases. 

A data consortium could use the information from member data to better inform bargaining. Members who are already getting good deals may be the best source of information about which packages or promotions are the best available for members who are getting worse deals. This need not require effort by members who are getting the best deals, just access to their data. This applies to subscriptions for services, and also to tuition and fees for universities, for payments for medical procedures, and to salary and benefits for employees. 

Many cities have transit riders' unions, which advocate on behalf of users of public transportation. Having access to member data could make these organizations more effective. For example, if they could analyze their members' aggregated travel patterns in detail, then they could advocate for specific investments to decrease crowding on the most congested routes or to extend routes to where the most members are traveling by other, more expensive, means. 

Data consortia could combine collective bargaining with aggregated member data to give consumers more power. For example, a data consortium whose membership includes of a significant fraction of the users of a service might develop an agreement among some of those users that if the price of the service rises above a specified amount, then those users will all unsubscribe from the service. Similarly, the data consortium might develop an agreement that some number of its members who are not using a service provider will switch to it if they can all get a specified bargain rate. It will be interesting to see whether and how aggregated data, analyzed on behalf of consumers, leads to collective action by consumers.

\section{Informed Living}
Aggregate data can improve quality of life. Searches and social media data can detect influenza outbreaks \cite{alessa18,sharpe16}. Using a combination of genealogy and genetic and medical data for the people of Iceland, deCODE was founded to develop medical diagnostics and drugs \cite{wired_icelanders}. This information has led to important discoveries in medicine and anthropology \cite{gudbjar15,ed15,ebenes18}. The nation of Bhutan collects data to regularly evaluate a measure called Gross National Happiness \cite{gnh_guide,gnh_book}, to understand the the state of well-being and the needs of its people.

For health, access to member data should allow data consortia to more accurately detect and even predict local outbreaks of illnesses. For example, access to emails could allow a consortium to identify which school or schools, and even which classes, each member's children attend. Combined with member search data or phone records, a consortium could identify which classes at a school are beginning to host an illness and also which nearby schools are likely to be infected next, through transmission via siblings. This would give parents a warning to find care options for kids or to perhaps keep them away from school for the most dangerous day or two. 

On a longer scale, access to member medical records and genetic data, similar to the Iceland database, and perhaps activity data from cell phones and fitness wearables and electronic receipts for food and activities, would give data consortia the ability to warn individual members about which conditions should concern them most, how to test for them, how to treat them, and, ideally, how to avoid them through lifestyle changes. Many health conditions in later life have their roots in habits or exposure in younger life; a health-focused data consortium could help members draw the connections and also help them distinguish the health issues that have the most impact on their lives from the noise. 

For happiness, sentiment analysis \cite{cambria13,ortony88,stevenson07,pang02} of member data, such as texts and emails and media consumed, can provide useful insights into the emotional states of members. In aggregate, this data can inform lifestyle decisions for individuals, for example it would be useful to know whether people who make a decision to walk or bike more instead of using a car for short trips become happier as a result. It would be interesting -- if perhaps somewhat controversial -- to find out whether adopting a dog or a cat makes a person happier.  

Sentiment analysis could also inform larger decisions. When selecting among jobs, selecting a town or city, or even selecting a neighborhood or building, people would find it useful to know where people tend to be happier, more optimistic, or more outgoing, and whether those things are changing for the better over time. In addition to providing this information to members, data consortia could sell sentiment analysis of aggregated member data to organizations. For retailers, it would be useful to find out whether customers find that they are happier after visiting the retailer's store or making purchases; the same information about competitors would also be valuable. For governments, it would be valuable to understand emotional well-being and its changes at a local level. It could inform policy decisions, in conjunction with economic measurements and projections. 

It will be interesting to see the impact of sentiment measurement on strategy. Will some governments work for the long-term well-being of their people, while others find that their power derives in some part from a certain level of disaffection among the people and so seek to perpetuate it? Will governments find it more efficient to improve the well-being in places that are lacking it or to encourage people to move to places where people are already living well and abandon unhappy places? Will local growth itself be a driver of well-being for some places but a harm for others? Similarly, will some companies attempt to make their customers fundamentally happy, while others seek to create a short-term emotional high, followed by a low that makes customers crave renewal, prompting another purchase?

\section{Discussion -- Challenges}
Data consortia -- organizations designed to use data on behalf of users -- have great potential, but they will face, and perhaps create, some novel challenges. Structurally, as a data consortium gains members, it gains statistical accuracy in its analysis and may also gain purchasing power. So a larger data consortium may offer more value to each member than a smaller one. As a result, one large data consortium with a natural monopoly may dominate for each purpose -- investing, spending, health, etc. -- and a large monopoly established for one purpose may find that it has a natural advantage in pursuing other purposes as well, simply because it already has access to data from, and established relationships with, so many members. 

However, where gains are shared among members, there may be pressure to keep data consortia small. In general, accuracy increases with the square root of the number of members, so there are diminishing returns to statistical analysis for each new member. If a data consortium can sell aggregate analysis of its members' data for a fixed price if it reaches a threshold of statistical significance, then members should only want enough other members to achieve that threshold. On the other hand, if members are also investors, then the returns scale with the number of members, while accuracy also increases and overhead costs per member decrease. So members should welcome members. For investing, though, at some scale, trades tend to move the market against the trader, perhaps counterbalancing the advantages of adding more members. 

Data consortium members may wish to be compensated directly for providing their data. If the analysis leads to decisions that have gains or losses, then the method outlined in Appendix \ref{value_data} can be applied. In other cases, members will have to come to agreements with consortia. Different members' data may have different value for different purposes. For example, a consortium that needs to draw inferences that apply to a larger population may have some segments of that population overrepresented among its members and other segments underrepresented. In general, this will make data from underrepresented-segment members more valuable, because it helps fill in gaps in analysis for the population as a whole. As another example, if a health-focused data consortium finds that a new drug is highly effective, then data from the first members who take the new drug and demonstrate its efficacy is very valuable, and the same is true for members who try out lifestyle changes that prove to be beneficial. In these cases, the value of a member's data is not necessarily known a priori; it can only be accurately assessed after the data are collected, outcomes are measured, and maybe even after the benefits accrue to other members.  

Biased \cite{baeza-yates18,oneil16} and false data \cite{lappas16,elmurngi17,mukherjee12} will be a challenge for data consortia. Suppose a data consortium sells aggregated information about the shopping habits of a small segment of people whose habits are of interest to a larger group of people, who wish to imitate them. The data consortium consists of members from the small segment, who are paid for their data. Then there is bias in the sense that the members are only from the subsegment of the small segment who are willing to give access to their data in exchange for payment. And there will be false data, because people who are not members of the small segment will attempt to falsely claim that they are, in order to become members of the data consortium. Also, sellers will have an incentive to sell to data consortium members on preferential terms or to create false identities and data streams for people in the small segment, make them members, and place receipts for their items in those false data streams. 

Similarly, suppose that a data consortium pays members more if they are part of some demographic segment, in order to create a balanced panel. Then members will have an incentive to falsely claim to be part of that demographic segment. Suppose that a data consortium examines email receipts of its members and buys stakes in companies that have increased receipts. Then companies have incentives to target members for marketing and even to email them false receipts. Just as there are myriad schemes to inflate ratings and sales numbers on many commercial websites, there will be efforts to feed data consortia false data.

\appendix

\section{Valuing Data} \label{value_data}
Shapley values \cite{shapley} offer a method to value individual contributions to a group effort: consider every possible order in which individuals could have agreed to include their contributions, and how much value the contribution would add to the previous contributions in each ordering, before any of the later contributions are added. Averaging this value over all orderings for a particular contributor assigns a value to their contribution. 

To apply Shapley values to a member's contribution of data to an investment, consider an ordering of members. Take the data from all members prior in the ordering to the member whose contribution is being evaluated. Evaluate the profit or loss that would have been realized using just that data instead of the full set of data from all members. Then add the data from the member whose contribution is being evaluated, and re-evaluate the profit or loss. The difference is the value assigned to the member's contribution for the ordering. Averaging this difference over all orderings yields the Shapley value for the member's contribution. It is computationally infeasible to compute this directly for even a moderate number of contributors. 

Instead, estimate by clustering and sampling.  Cluster the members according to their data. For each cluster, estimate the Shapley value for a sample of users, and assign the sample average to all members in the cluster. For each sample member, using sampling over the orderings. Use stratified sampling to estimate the average difference in gain or loss due to adding the member's data to data from each number of members. For example, begin with all data and remove data from other members selected at random until there is a difference in the trading decision. Begin by removing half the remaining members' data. If there is a difference in the trading decision, then add back half the removed members; otherwise also remove half the remaining members, performing a binary search for the number of members needed for the member being evaluated to make a difference in the outcome. Repeat this binary search procedure for different randomly selected sets to estimate the distribution of number of members that must be removed to make a difference in the outcome. 

A basic outline of the process for using member data to inform investing is:

\begin{enumerate}
\item \textit{Data}. Fetch data, check, and clean and/or normalize.
\item \textit{Model}. Apply a model to the data. The model may be trained on the data to set parameters or have weights set to make the demographics or other categories of members reflect a larger population. 
\item \textit{Score}. Applying the model to the data produces a score.
\item \textit{Check Statistical Significance}. If the thresholded score is statistically significant, then trade on the signal. Statistical significance depends on both the score and the amount of data that produces it. 
\item \textit{Realize a Profit or Loss}. If there is trade, evaluate its profit or loss (relative to not trading). With no trade, the relative profit or loss is zero.
\end{enumerate}

To determine Shapley values, apply this process to data from subsets of members as outlined previously, using the model to produce a score and checking for statistical significance to produce the trading decision that would have been made if the subset was the entire data set, and, if a trade would have been executed, then use historical financial data to evaluate its profit or loss. In this way, it is possible to assess changes in profit or loss due to adding a member's data to data from a subset of members. 

Note that a member's data can change the decision to trade in several ways. The data might affect the model weights, whether they are determined by training on the data or used to make the members a balanced panel that represents a larger population. The member's data directly affects the score, since the data is the input and the score is the output of the model. Member data also affects statistical significance, so adding a member's data may change the decision to trade, even if their data influences the score in the opposite direction.

\bibliographystyle{ACM-Reference-Format}
\bibliography{bax.bib}

\end{document}